\documentclass[pra,preprint,tightenlines,showpacs,twocolumn,10pt]{revtex4} 

\usepackage{epsfig} 
\usepackage{graphicx}
\usepackage{amsmath, amsfonts, amssymb, bm}
\usepackage{braket}
\usepackage{color}
\usepackage[utf8]{inputenc} 
\usepackage{subfigure}
\usepackage{soul,ulem,cancel}
\begin{document}  










\title{ Photo ionization of an atom passed through a diffraction grating }
\author{ S. F. Zhang$^1$ , B. Najjari$^1$, X. Ma$^1$\footnote{x.ma@impcas.ac.cn} and A. B. Voitkiv$^2$\footnote{alexander.voitkiv@tp1.uni-duesseldorf.de} }  
\affiliation{ 
$^1$ Institute of Modern Physics, Chinese Academy of Sciences, Lanzhou 730000, China \\ 
$^2$ Institute for Theoretical Physics I, Heinrich-Heine-University of D\"usseldorf, 
Universit\"atsstrasse 1, 40225 D\"usseldorf, Germany } 

\date{\today} 

 
\begin{abstract} 
We consider photo ionization of an atom which, due to passing through 
a diffraction grating, is prepared in a multi-site state possessing 
a periodic space structure with alternating maxima and minima. 
It has been found that this process qualitatively differs from photo ionization of 
a 'normal' atom. In particular, the spectra of emitted electrons and 
recoil ions in this process 
display clear one- and two-particle interference effects. Moreover, 
there are also striking differences between the momentum distributions 
of these particles, which no longer mirror each other. 
The origin of all these features is discussed in detail. 
It is also shown that the information about the diffraction 
grating, which is encoded in the multi-site state of the atom,   
can be fully decoded by exploring the spectra of recoil ions 
whereas the photo-electron spectra contain this information 
only partially.   

\end{abstract} 

\maketitle  

\section{ Introduction } 

Photo ionization of atomic systems 
belongs to the most 'popular' processes 
studied by atomic physics.
Since the discovery of the photoelectric effect  
numerous studies on photo ionization of atoms, ions and molecules 
have not only greatly improved our knowledge of their structure 
but also substantially contributed to the understanding of 
the basic laws of quantum physics. Modern experiments on 
photo ionization of atoms and simple molecules 
performed using COLTRIMS techniques \cite{coltrims} 
can 
provide a complete information on  
the momenta of all reaction fragments. 

Break-up of a hydrogen-like atomic system 
due to absorption of a photon is the most basic 
(and simple) photo ionization process.  
It has been scrutinized in numerous 
studies and its quantum dynamics has been understood 
in very great detail.

The process of photo ionization of molecules is more complex and 
its description and understanding are substantially more complicated. 
The basic prototype of this process is represented by 
photo break-up of a system, which consists of a single electron and two nuclei 
(e.g., H$^+_2$) forming a bound state due to the interactions between them.    
One of interesting effects, which can arise in photo ionization of 
such a system, is interference (observed e.g. in the electron emission pattern). 

The origin of this interference can be qualitatively understood by regarding 
the state of the electron in the molecule, where it 
simultaneously orbits around two nuclei, as a superposition 
of two atomic-like electron states centered on the nuclei. 
Within such a simple picture the electron, by absorbing a photon,  
is launched simultaneously from the two sites of the molecule. 
As a result, two undistinguishable reaction pathways arise 
that leads to interference.   

\vspace{0.25cm} 

In an atom, where an electron orbits around only one nucleus, 
multi-site electron states, similar to those in a molecule, 
of course cannot exist. 
Therefore, in photo ionization of an atom 
there are no interference effects, which would be caused 
by a multi-site structure of the electron state. 
Indeed, when the wave function of 
a free atom is represented (as usual) by 
\begin{eqnarray}  
\Psi({\bm R}, {\bm r}, t) \sim e^{ i ( {\bm P} \cdot {\bm R} - E_p t) }  
\times  \varphi({\bm r}) \exp(- i \varepsilon t ), 
\label{normal_atom}
\end{eqnarray} 
where the plane-wave refers to the center-of-mass motion of 
the atom and the rest describes the internal atomic degrees of freedom, 
such interference effects are naturally absent. 

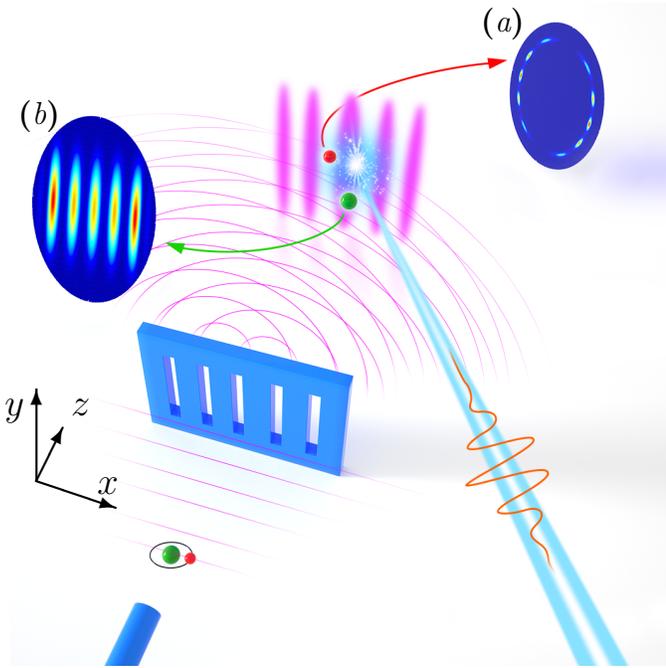
\begin{figure}[ht]
\center
\includegraphics[width=0.49\textwidth]{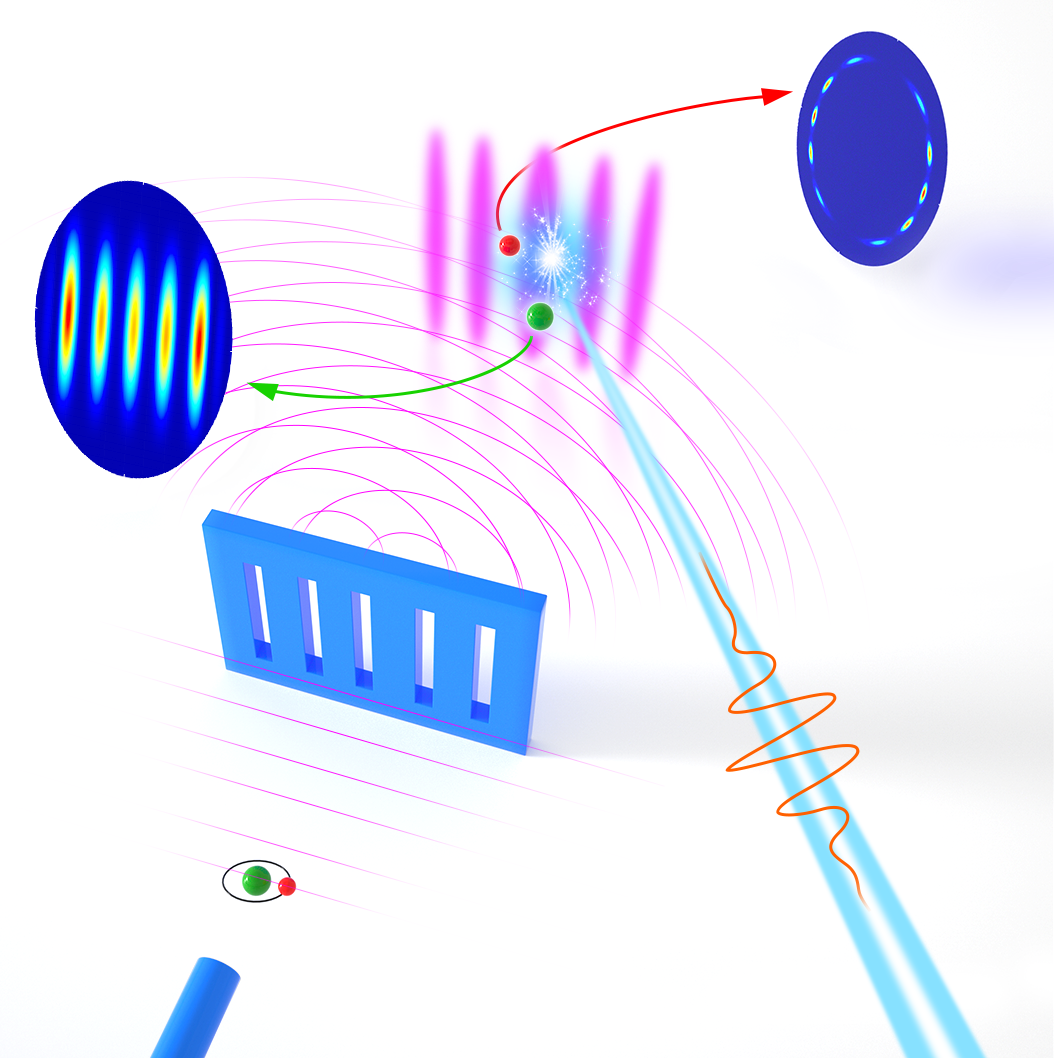}
\put(-75,240){\color{black} {\fontsize{12}{8} \selectfont {\bf ({\it a})} }}   
\put(-250,205){\color{black} {\fontsize{12}{8} \selectfont {\bf ({\it b})} }}   
\thicklines
\put(-240,70){\color{black}\vector(0,3){35}}
\put(-240,70){\color{black} \vector(3,-1){30}}
\put(-240,70){\color{black} \vector(1,2){10}}
\put(-255,95){\color{black} {\fontsize{14}{48} \selectfont $y$} }
\put(-230,95){\color{black} {\fontsize{14}{48} \selectfont $z$} }
\put(-220,65){\color{black} {\fontsize{14}{48} \selectfont $x$} }
 \vspace{.2cm}
\caption[caption]{ A sketch of preparing and photo ionizing 
a 'multi-site' atom. 
An atom with a well-defined initial momentum after passing through a diffraction grating 
(blue screen with slits) finds itself in a multi-site  
state (purple fringes). 
In the interaction region the atom is ionized by a light beam (marked with cyan and orange). 
The momenta of the emitted electron (red sphere) and the recoil ion (green sphere) 
can then be measured using two dimensional position-sensitive detectors, 
labeled by (a) and (b), respectively.}
\label{figillustration} 
\end{figure}

However, even a single atom can be prepared in a state possessing  
a multi-site structure. This can be done, for instance, by letting an atom to pass 
through a multi-slit screen (a diffraction grating) as illustrated in Fig.\ref {figillustration}. 
After the passage, the atomic wave function acquires a periodic 
space structure with alternating maxima and minima in its absolute value 
(depicted in the figure by pink fringes). This, to some extend, can be viewed 
as 'splitting' the atom into a set of identical atomic 'copies' 
with equidistant separation between them.   
Such a multi-site atomic structure (termed in \cite{we-qdg} 'quantum grating'), 
is organized based on only one 'real' nucleus 
(and, in the simplest case, only one 'real' electron). Hence, such 
a 'multi-site' atom qualitatively differs from an object like e.g. H$^+_2$, 
where a two-site structure of the electron state is created by the presence of 
two 'real' nucleus. 

Therefore, there are all grounds to expect that the process of photo ionization 
of a 'multi-site' atom will strongly differ from photo ionization 
of not only a 'normal' atom but also a molecule. 
It is thus the goal of the present article 
to theoretically explore photo ionization of 
such a 'multi-site' 
atom taking, as its simplest representation, 
a hydrogen-like atomic system.  

The atomic units are used throughout the text except otherwise stated. 

\section{ General consideration }  

Let an atom (or an atomic ion) with a momentum ${\bf P}_i$ ($P_{i, x} = 0, P_{i,y} = 0, P_{i, z} = P_i$) 
pass through a diffraction grating. We suppose that the grating is located in the $(x$-$y)$--plane 
(i.e., perpendicular to ${\bf P}_i$) and consists of $N_0$ equal-size slits placed along the $x$-direction 
(see Fig.~\ref{figillustration}). The dimensions of the slit 
along the $x$- and $y$- directions are $a$ and $b$, respectively.  
The separation between the neighbor slits along the $x$-direction 
(the grating constant) is $d$. 

After the passage through the grating the wave function of the atom at large distances 
can be obtained using the Huygens-Fresnel principle (see e.g. \cite{H-FP}) 
and is very well approximated by 
\begin{eqnarray} 
\Psi_{i}({\bm R},{\bm r}, t) & \propto & { e^{iDP_i}} 
\, \, \, e^{ i \big( { P_i Z -  E_i t } \big) } \times \notag\\
&& { \sin\left( {\alpha X }\right) \over \alpha X}  \, 
{ \sin\left( N_0 {\gamma X }\right) \over \sin\left(\gamma X\right) } \, 
{ \sin\left( {\beta Y }\right) \over \beta Y} \varphi_i({\bm r}) .   
\label{QG-WP} 
\end{eqnarray}
Here, ${\bm R} = (X,Y,Z)$ is the position vector of the atomic center-of-mass 
with respect to the origin, ${\bm r } $ is the electron position vector with respect to the atomic nucleus,   
$ \varphi_i({\bm r})$ is the initial internal state of the atom with an energy $ \varepsilon_i $, 
and $E_i = \frac{P_i^2 }{ 2M_A} + \varepsilon_i$ is the total energy of the atom with 
$ M_A $ being its total mass, $ \alpha = \frac{ P_i \, a}{ 2 \, D } $, $ \beta = \frac{ P_i \, b}{ 2 \, D } $ 
and $ \gamma = \frac{ P_i \, d}{ 2 \, D }$. 
As the origin of our coordinate system we have taken the center of the interaction region 
(where the atom and photon beams cross) whose linear dimensions are assumed to be much less 
than the distance $D$ between this region and the diffraction grating. 

In the interaction region the atom is irradiated by an electromagnetic plane wave 
with a (central) frequency $\omega_\kappa$, momentum $ {\bm \kappa} $ and polarization vector 
$ {\bm \epsilon}_{\bm \kappa} $  ($ {\bm \kappa} \cdot {\bm \epsilon}_{\bm \kappa} = 0$). 
Absorption of a photon results in electron emission. Assuming that the momenta of the emitted 
electron and recoil ion can be measured with high accuracy, we take the final state of the atom 
as the product of a plane wave, which describes the motion of its center-of-mass, 
and an internal (continuum) state $ \varphi_{{\bm k}_e}({\bm r})$ with an energy 
$ \varepsilon_{k_e} =  {k_e}^2/2 $ where ${{\bm k}_e}$ 
is the momentum of the emitted electron with respect to the atomic nucleus.

After a (somewhat lengthy) calculation we obtain that the fully differential cross section 
for photo ionization is given by 
\begin{widetext}  
\begin{eqnarray}
\frac{ d\sigma_{fi} }{ d^3 {\bm P}_{rec}\; d^3 {{\bm p}_e} }
\propto && { 1\over \omega_\kappa} { \delta(E_f -E_i - \omega_\kappa)} 
{ \delta(P_i-P_{f,z} +\kappa_z)}  \notag\\
&&\times\bigg|\Big<
\varphi_{{\bf k}_e}\Big|  e^{i\boldsymbol\kappa\cdot{\bf r}} 
\big( \boldsymbol \epsilon_\kappa \cdot \hat{\bf p }_r \big) \Big|  
\varphi_i\Big>\bigg|^2 \notag\\
&&\times 
\bigg| F_\beta(P_{f,y} - \kappa_y) \sum_n F_\alpha(P_{f,x} - \kappa_x + n P_i d/D)  \bigg|^2 . 
\label{cross-section1}
\end{eqnarray} 
\end{widetext} 
In this expression ${\bm P}_f = (P_{f,x},P_{f,y},P_{f,z})$ 
and $E_f = {P_f}^{2} /2 M_A  + {k_e}^2/2$ are the final total momentum and energy, respectively, 
of the atomic system (the electron + residual ion), and 
$ {\bm P}_{rec} $ and ${\bm p}_e \simeq {\bm k}_e + {\bm v}$, where 
${\bm v} = {\bm P}_i/M_A $, are the momenta of the recoil ion and the emitted electron, 
respectively, ($ {\bm P}_{rec} + {\bm p}_e = {\bm P }_f $); 
all these momenta and the energy refer to the laboratory frame. 
$\hat{\bm p }_{\bm r} $ is the momentum operator acting on the internal states. 
The sum in (\ref{cross-section1}) arises due to the periodic structure 
of the state (\ref{QG-WP}) running over the number of the 'localization sites' of the state (this number is equal to the number of slits of the diffraction grating). 

The two delta-functions in (\ref{cross-section1}) ensure the momentum conservation 
along the $z$-axis, given by $P_{rec,z} = P_i + \kappa_z - p_{e,z}$, 
and the energy conservation, which reads $ ({\bf p}_{e}-{\bf v})^2/2 = 
\omega_\kappa - {\bm v} \cdot {\bm \kappa}  + \varepsilon_i$, where   
the term $ {\bm v}\cdot{\bm \kappa}$ accounts for the (non-relativistic) 
Doppler shift of the photon frequency in the rest frame of the atom 
(note that it is quite small and can normally be neglected). 

Had we considered photo ionization of a 'normal' atom, 
the corresponding fully differential cross section would involve  
two more delta-functions expressing momentum conservation 
along the $x$- and $y$- directions. In the case of a 'multi-site' atom, however, 
they are replaced by the terms in the last line of Exp. (\ref{cross-section1}), 
where the functions $F_\eta $ ($ \eta = \alpha, \beta$) are given by    
\begin{eqnarray}
F_\eta(\zeta)  =&& \int_{-L/2}^{+L/2} \!\! dl\; \;e^{i \, \zeta \, l}   \;\;  
{ \sin\left( {\eta \, l }\right) \over  \eta \, l}. 
\label{F-functions}
\end{eqnarray} 
Here, $L$ is the length of the interaction region (for simplicity 
we assume that it is the same for both 
$x$- and $y$- directions).   
If $L \gg 4 \, \pi \, \eta $,  
the function $F_\eta(\zeta)$ takes on very simple form: 
$ F_\eta(\zeta) = \pi/\eta $ at $ | \zeta | < \eta $, 
$ F_\eta(\zeta) = \pi/(2\, \eta) $ at $ | \zeta | = \eta $, 
and $ F_\eta(\zeta) = 0 $ at $ | \zeta | > \eta $. 

The conditions $ L \gg 4 \, \pi \, \alpha $ 
\big($ L \frac{ P_i a }{ D }  \gg 2 \pi $\big) and 
$ L \gg 4 \, \pi \, \beta $ 
\big($ L \frac{ P_i b }{ D }  \gg 2 \pi $\big)  
hold for a very broad parameter range and 
in what follows we shall assume that they are fulfilled.  
In such a case Exp. (\ref{cross-section1}) predicts that 
the momenta of the reaction fragments must fall into the momentum bands 
defined by 
$ | P_{rec,x} + p_{e,x} - \kappa_x + n \, P_i \, \frac{ d }{ D} | \leq P_i \, \frac{ a }{ 2 D} $ 
and 
$ | P_{rec,y} + p_{e,y} - \kappa_y | \leq P_i \, \frac{ b }{ 2 D} $. 
These requirements, together with the momentum conservation 
along the $z$-axis and the energy conservation,  
determine the photo ionization kinematics. 

Note that the absence of the delta-functions 
for the momentum balance in the ($x$-$y$)--plane 
is a direct consequence of regarding the diffraction grating 
as an external field acting on the atom. 
The total momentum of the system consisting of the atom, 
the grating and the photon is conserved 
leading to a strong (momentum) entanglement between the first two 
(since the photon momentum can normally be neglected).  
However, since the grating's degrees of freedom are not included 
in the consideration,  
the momentum conservation is sacrificed. In contrast, 
the energy conservation remains since the macroscopic 
grating (which was initially at rest), due to its enormous mass, 
does not participate in the energy exchange with the atom. 
  
\section{ Results and Discussion }  

\subsection{ Spectra of electrons and recoil ions } 

\begin{figure}[h!]
\center
\includegraphics[width=0.51\textwidth]{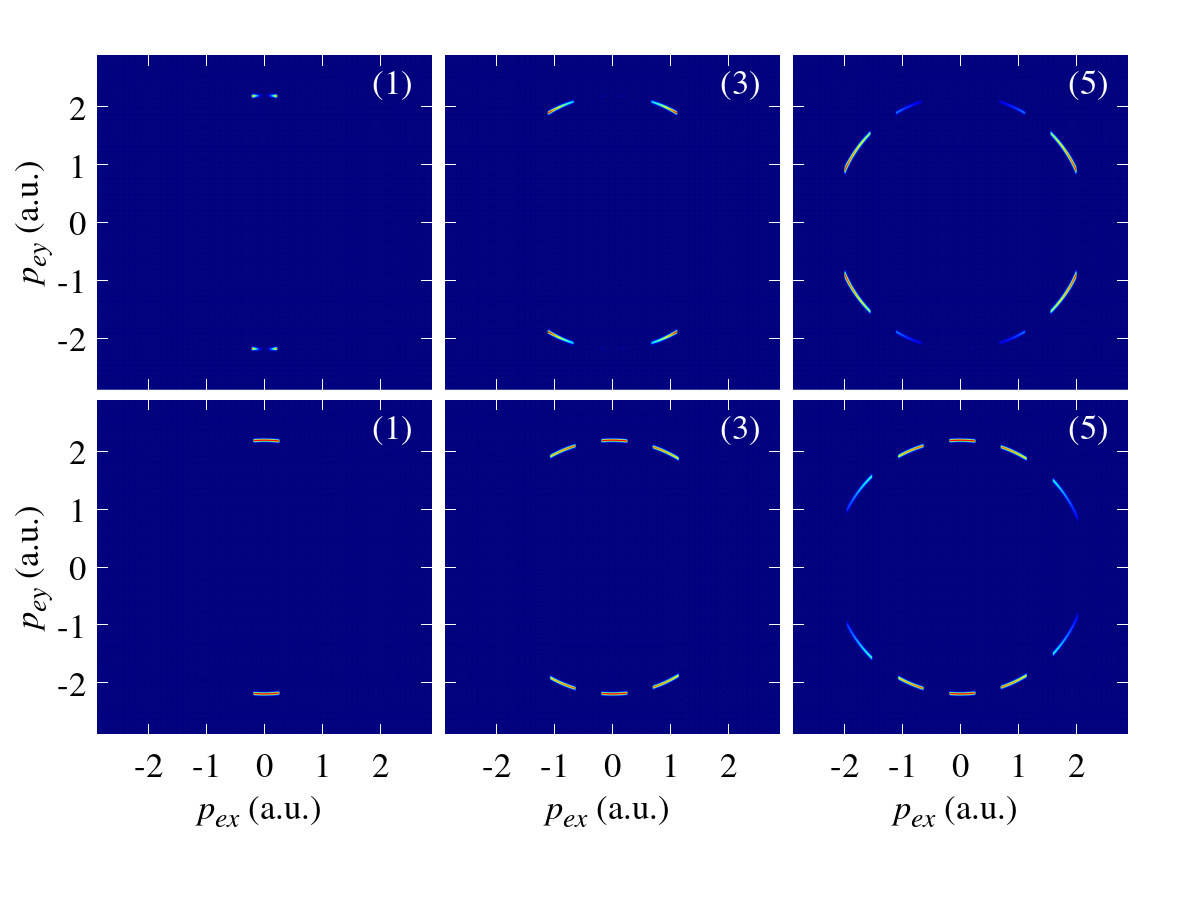}
\vspace{-1.1cm}
\caption[caption]{ The cross section 
$ \frac{ d \sigma }{ d^3 {\bm p}_{e} \, \, dP_{rec,x}}$ 
for photo breakup of He$^+$(1s) by a linearly plane wave  
with central frequency $\omega_\kappa = 120$ eV and bandwidth 
$ \Delta \omega_\kappa = 1 $ eV. 
The multi-site 
state of He$^+$(1s) was formed by passing with the initial 
momentum $P_i = 8\times 10^2 $ a.u. ($v \approx 0.1$ a.u.) 
through a diffraction grating with $a = d/2 = 100 $ $\mu m$.  
The distance $D$ between the grating and the interaction region is $18$ cm. 
The cross section is plotted as a function of 
$p_{e,x}$ and $p_{e,y}$ at fixed $p_{e,z} = v \approx 0.1$ a.u. and $ P_{rec,x}=0 $ . 
The upper and lower panels correspond to polarization 
along the $x$- and $y$- axis, respectively, 
the number of slits is marked in the upper-right corner of each panel. }
\label{figure-electron}  
\end{figure} 

Fig. \ref{figure-electron} shows spectra of electrons emitted from 
multi-site 
He$^+$(1s) ions due to absorption of    
photons from a linearly polarized radiation  
with a central frequency $ \omega_\kappa = 120 $ eV and a bandwidth 
$\Delta \omega_\kappa = 1$ eV.  
The multi-site 
state of the He$^+$(1s) is formed by passing 
with initial momentum $P_i = 800 $ a.u. through a diffraction grating 
($a = d/2 = b/2 = 100 \;\mu m $). 
The distance between the grating and the interaction region 
and the linear dimension of the latter are $D = 18 \; cm$ and 
$L = 1$ mm, respectively.  
The spectra are represented by 
the cross section $ \frac{ d \sigma }{ d^3 {\bm p}_{e} \, \, dP_{rec,x}}$ 
and are given as a function of the electron momentum 
components $ p_{e,x} $ and $ p_{e,y} $ at $ p_{e,z} = v \approx 0.1$ a.u. 
and $P_{rec,x} = 0$. The upper and lower panels of Fig. 
\ref{figure-electron} display results for the absorption of 
photons polarized along the $x$- and the $y$- axis, respectively. 

Had we considered photo breakup of the 'normal' He$^+$(1s) 
(assuming that $ a \to \infty $ and $ b \to \infty$ and keeping 
the other parameters as they are given in Fig. \ref{figure-electron})   
the corresponding cross section $ \frac{ d \sigma }{ d^3 {\bm p}_{e} \, \, dP_{rec,x}}$ 
would be represented by just two short lines of infinitesimally small width 
along the $x$-axis (their size along the $y$-axis is proportional to $\Delta \omega_\kappa $), which are  
located at $(p_{e, x}; \, p_{e, y}) = (0; \pm \sqrt{ 2(\omega_\kappa - |\varepsilon_i| ) } $.    
Moreover, a similar result would also hold for ionization of a molecule. 

In contrast, the electron spectra in Fig. \ref{figure-electron} 
possess a clear interference structure originating 
in the coherent contributions of different sites  
of the state (\ref{QG-WP}) to the photo breakup process. The main 
features of this structure can be qualitatively 
understood by noting the following points.   
First, because of the energy conservation the photo electron 
spectra in the ($p_{e,x}$-$p_{e,y}$)--plane 
can only be located on a ring with the (mean) radius 
$ p^0_{\perp} = \sqrt{ 2( \omega_\kappa - |\varepsilon_i|) - (p_{e,z}-v)^2 } = 
\sqrt{ 2( \omega_\kappa - |\varepsilon_i|) }$;  
due to a finite radiation bandwidth $\Delta \omega_\kappa $ 
the ring has a width $ \Delta p^0_{\perp} \approx \Delta \omega_\kappa/p^0_{\perp} $. Second, 
the dipole selection rules (the dipole approximation is valid 
since the photon momentum is negligible) 
'encourage' the emitted electron to move along the photon polarization axis. 
Third, and what makes the crucial difference compared to the 'standard' photo ionization, 
is that in the momentum balance along the $x$-axis, which is determined by the function $F_{\alpha}$ 
(see the discussion after Exp. (\ref{F-functions})), the strict unambiguous relation  
$p_{e,x} = - P_{rec,x}$ ($ = 0 $ a.u.) is now replaced by the much milder condition of 
prohibiting the emitted electron to fall outside the momentum bands  
$ | p_{e,x} + n \, P_i \, \frac{ d }{ D} | \leq P_i \, \frac{ a }{ 2 D} $ 
($n = 0, \pm 1, \pm 2$,...). 

Using the above three points one can qualitatively 
(and even quantitatively) explain: 
i) the positions of the maxima in the spectra 
(including the double-splitting of the single maximum in the upper-left panel 
of Fig. \ref{figure-electron}), 
ii) how the spectra change with the number of slits 
and also iii) the appearance of the 'saturation' in the spectrum pattern 
with increasing the number of slits $N_0$  when neither the number of 
the spectrum maxima nor their positions and shape  
noticeably change with a further increase of $N_0$:  
in particular, in the case under consideration 
the pattern becomes essentially independent of $N_0$ 
starting already with $ N_0 = 5 $ .  

\begin{figure}[h!]
\center
\includegraphics[width=0.51\textwidth]{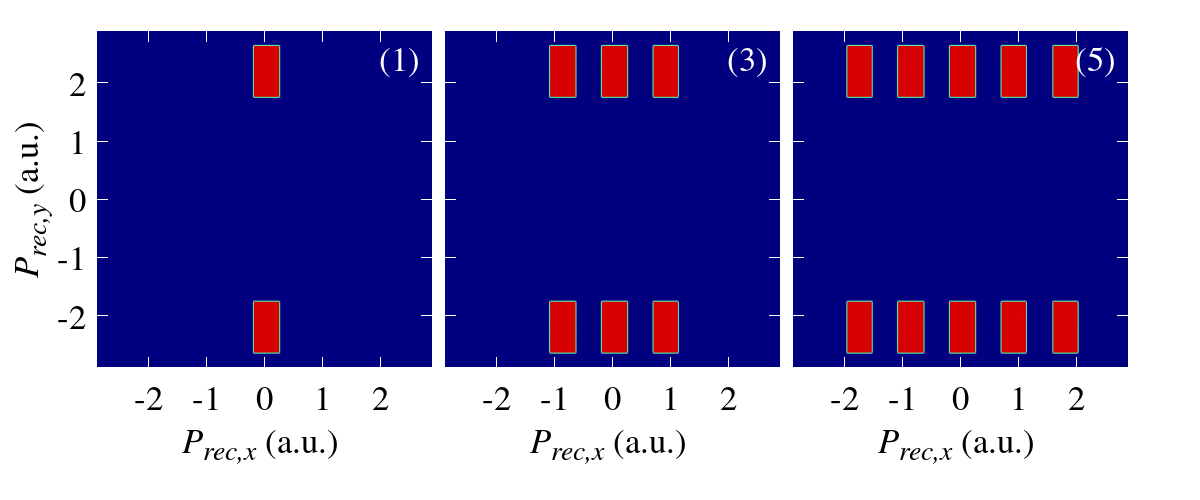} 
\vspace{-0.95cm}
\caption[caption]{ The cross section $ \frac{ d \sigma }{ d^3 {\bm P}_{rec} \, \, dp_{e,x}}$ for 
the same process as in Fig. (\ref{figure-electron}).  
The cross section is plotted as a function of 
$P_{rec,x}$ and $P_{rec,y}$ at fixed $P_{rec,z} = P_i - v$ and $p_{e,x}=0$ a.u. 
The radiation is polarized along the $y$-axis,  
the number of slits is marked in the upper-right corner of each panel. }
\label{figure-rec-ion}  
\end{figure} 

Fig.~\ref{figure-rec-ion} displays spectra of the recoil ions 
represented by the cross section 
$ \frac{ d \sigma }{ d^3 {\bm P}_{rec} \, \, dp_{e,x}}$.  
It was calculated for the same, as in Fig.~\ref{figure-electron}, 
values of the initial momentum $P_i$, the distance 
$D$, the size of the interaction region $L$ and the parameters of 
the radiation and the diffraction grating. 
The recoil ion spectrum 
is given as a function of the recoil ion momentum components 
$ P_{rec,x} $ and $ P_{rec,y} $ at $ P_{rec,z} = P_i - v \approx 800$ a.u. and 
$p_{e,x} = 0$ a.u. 

This choice of the cross section for the recoil ions 
(and of the latter two momentum components)  
was made in order to 'mirror' the conditions for the electron emission spectra 
in Fig.~\ref{figure-electron}. 
However, when the radiation is polarized along the $x$-axis 
there is no 'mirroring' between the electron and recoil ion spectra at all 
(since the latter simply vanishes at $p_{e,x} = 0$).  
If the radiation is polarized along the $y$-axis, 
there is indeed a certain correspondence between these spectra. 
Yet, in contrast to photo ionization of a 'normal' atom (or molecule) in which 
the momentum spectra of electrons and recoil ions 
would be very strongly correlated exactly mirroring each other,  
this correspondence does not imply such a strong correlation 
even in the single-slit case ($N_0 = 1$)  
and further rapidly diminishes when the number of slits increases.     
  
The positions of the maxima in Fig.~\ref{figure-rec-ion} are determined 
by the conditions \\ $ | P_{rec,x} + n \, P_i \, \frac{ d }{ D} | \leq P_i \, \frac{ a }{ 2 D} $  
and $ | P_{rec,y} + p^0_{e,y} | \leq P_i \, \frac{ b }{ 2 D} $. 
The value of $p^0_{e,y}$ is determined by the energy conservation,   
$ [p^2_{e,x} + (p^0_{e,y})^2  + (p_{e,z}-v)^2]/2 =  \omega_\kappa - |\varepsilon_i| $, 
where now $p_{e,x} = 0$, $ p_{e,z} = P_i - P_{rec, z} = v $ and, hence, 
$p^0_{e,y} = \pm \sqrt{ 2(\omega_\kappa - |\varepsilon_i|) } \approx \pm \, \, 2.2$ a.u.  
Due to the radiation bandwidth $\Delta \omega_\kappa $ there is 
a momentum uncertainty 
$ \Delta p^0_{e,y} \approx \pm \Delta \omega_\kappa/(2 \, p^0_{e,y}) \approx \pm 0.01$ a.u.  
(note that $ P_i \, \frac{ a }{ 2 D} =  P_i \, \frac{ b }{ 2 D} \approx 0.22$ a.u.). 

These conditions, in particular, show that the number of the maxima 
in the recoil ion spectra is simply equal to the number of slits $N_0$ 
and, unlike the electron spectra, 
no 'saturation' in the pattern of the recoil cross section occurs 
when this number grows. The reason for this is 
that -- because of the huge difference between the electron and nucleon masses -- 
the recoil ion momenta $ P_{rec,x} $ and $ P_{rec,y} $ do not enter   
the expression for the energy conservation in photoeffect, 
$ ({\bf p}_{e}-{\bf v})^2/2 = 
\omega_\kappa  - |\varepsilon_i| $, 
and the possible values of $ P_{rec,x} $ are unrestricted by the energy constraints. 

\begin{figure}[h!]
\center
\includegraphics[width=0.51\textwidth]{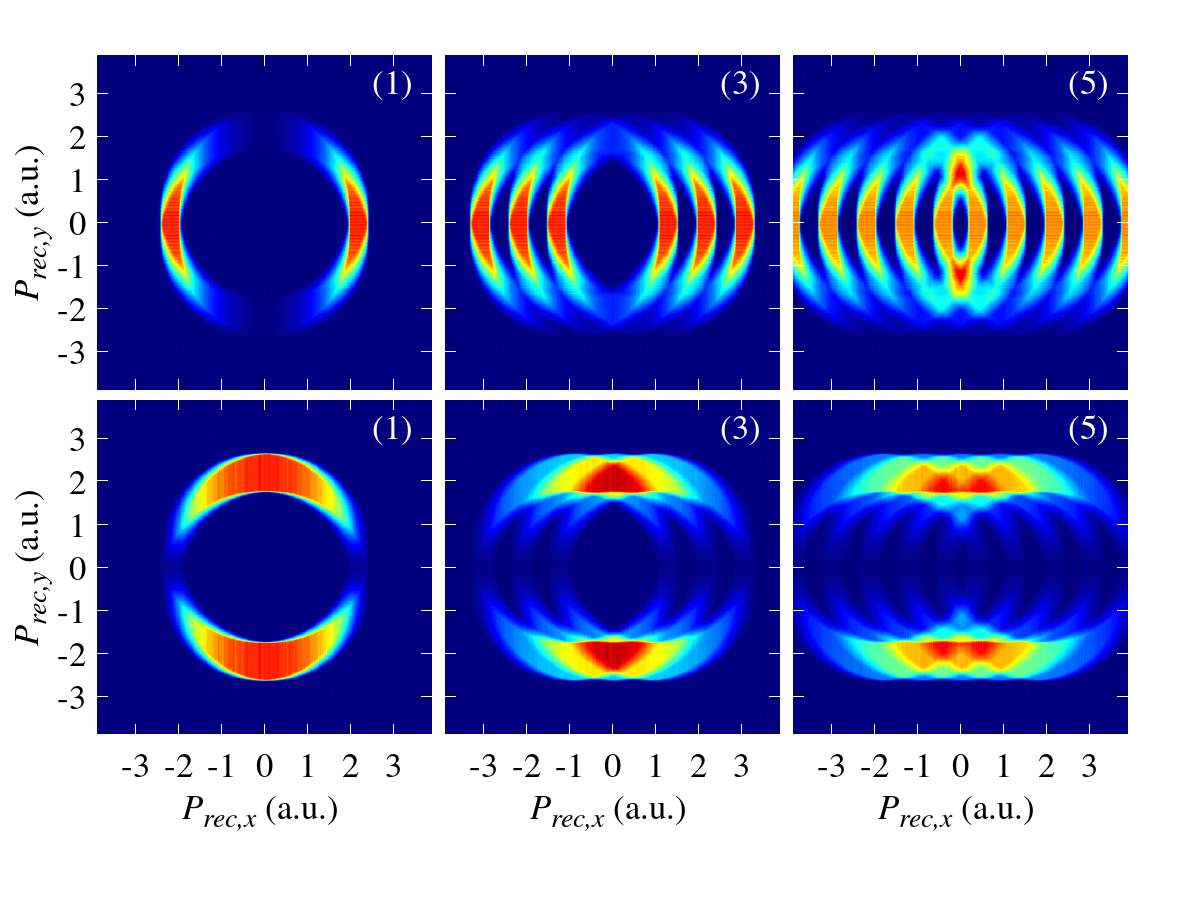} 
\vspace{-0.95cm}
\caption[caption]{ The cross section $ \frac{ d \sigma }{ d^3 {\bm P}_{rec} } $ 
given as a function of $P_{rec,x}$ and $P_{rec,y}$ at a fixed $P_{rec,z} = P_i - v$. 
The parameters $P_i$, $a$, $b$, $D$, $L$, $\omega_\kappa$ and $\Delta \omega_\kappa$
are the same as in Figs. \ref{figure-electron}--\ref{figure-rec-ion}. 
The upper and lower panels show results for 
the radiation linearly polarized along the $x$- and $y$- axis, respectively. 
The number of slits is marked in the upper-right corner of each panel. }
\label{figure-rec-ion-int}  
\end{figure} 
 
Let us, as the last examples, consider the cross sections 
$ \frac{ d \sigma }{ d^3 {\bm p}_{e} } $ and $ \frac{ d \sigma }{ d^3 {\bm P}_{rec} } $. 
A simple analysis of the fully differential cross section (\ref{cross-section1}) 
shows that after its integration over all possible states of the recoil ion 
the resulting electron momentum spectrum acquires exactly the same shape  
as in the case of photo breakup of a 'normal' atom (or ion). Namely, 
the electron momenta in the three dimensions are located on a sphere with the (mean) radius 
$ \sqrt{ 2(\omega_\kappa - |\varepsilon_i|) } $ 
centered at $(0; 0; v)$; the sphere is broadened due to the radiation bandwidth and  
the intensity distribution on it depends on the polarization of the radiation. 
Fixing a value of $p_{e,z}$ we obtain that in the ($x$-$y$)--plane 
the electron momenta are located on a (corresponding) ring.   

The shape of the cross section $ \frac{ d \sigma }{ d^3 {\bm P}_{rec} } $ 
is more interesting, qualitatively differing from that 
for photo ionization of a 'normal' atom (the latter just exactly mirrors the corresponding 
electron spectrum). This cross section is shown in Fig. \ref{figure-rec-ion-int}, 
where it is plotted as 
a function of $P_{rec,x}$ and $P_{rec,y}$ at a fixed $P_{rec,z} = P_i - v$. 
According to this figure the momenta of the recoil ions in the ($x$-$y$)--plane 
are located on (multiple) rings with the mean radius of 
$ \sqrt{ 2(\omega_\kappa - |\varepsilon_i|) }  \approx 2.2$ a.u. centered 
at the points ($ P_{rec,x} = n \, P_i \, \frac{ d }{ D }; \, P_{rec,y} = 0 $) 
$\approx$ ($ 0.89 \, n; \, 0 $) with $n = 0, \pm 1, ...$ . 
The rings have a width due to a finite radiation bandwidth $\Delta \omega_\kappa$ 
(this contributes $ \approx 0.02 $ a.u. to the width of the rings)   
and the momentum bandwidths in the $x$- and $y$- directions given by 
$ \Delta P_x = P_i \, \frac{ a }{ D} \approx 0.44$ a.u. 
and $ \Delta P_y =  P_i \, \frac{ b }{ D} \approx 0.44$ a.u., respectively. 
Thus, at $N_0 > 1$ striking qualitative differences arise 
between the 'individual' momentum distributions of the emitted electrons and recoil ions.  

The only fundamental parameter making a drastic difference 
between the electron and the nucleus of the He$^+$ is their huge mass difference. 
Therefore, it is obvious that it is exactly this point where 
the lack of symmetry between the momentum spectra of the photo reaction 
fragments originates.  
Even so this point is of course present also for a 'normal' atom 
(where nevertheless the symmetry between the momentum spectra is present), 
here its effect is qualitatively different due to a specific state of the atom. 
Indeed, the atom initially incident along the $z$-direction, 
due to scattering on the diffraction grating, 
falls into the allowed momentum bands 
$ \big( n \, P_i \, \frac{ d }{ D } -  P_i \, \frac{ a }{ 2 D } 
\leq P_x \leq  n \, P_i \, \frac{ d }{ D } +  P_i \, \frac{ a }{ 2 D }; 
-  P_i \, \frac{ b }{ 2 D } \leq P_y \leq  P_i \, \frac{ b }{ 2 D } \big) $  
and, because the electron is much lighter than the nucleus, it is mainly 
the latter which participates in the momentum exchange with the grating. 

\subsection{ Photo ionization of a 'multi-site' atom  
as one- and two-particle interference phenomena } 

As we have seen,  if the diffraction grating contains more than one slit 
all the above considered cross sections (except  
$ \frac{ d \sigma }{ d^3 {\bm p}_{e} } $) display pronounced 
interference patterns. The interference effects are caused by 
a periodic multi-site space structure 
of the state of an atom passed through a diffraction grating.  
Because of that the atom (atomic ion) in the process of photo ionization 
effectively behaves as a set of equidistant coherent absorbers of radiation 
that -- according to the superposition principle -- gives rise to interference. 

However, the interference structures in the electron emission spectrum 
disappear if it is obtained by integrating over all final states 
of the recoil ion. In other words, interference is absent 
for the electron taken individually and appears   
only if the detection of the electron is accompanied 
by a coincidence measurement of the recoil ion. 
Thus, the emission pattern in Fig. \ref{figure-electron}    
represents in fact two-particle interference phenomena 
arising provided the superposition principle and quantum entanglement 
of the electron and recoil ion 'act' together  
(two-particle interference phenomena are considered e.g. 
in \cite{2-particle-2}-\cite{2-particle-4}).

In contrast, in the recoil ion spectra the interference structures 
are present, no matter whether the emitted electron is detected or not. 
Thus, in this case interference remains even if the recoil ion 
is taken individually belonging thus to the class 
of one-particle interference phenomena.    

\subsection{ Photo ionization of a 'multi-site' atom 
as the process of extracting information } 

The process of photo ionization (breakup) can also be viewed from 
a somewhat different perspective: namely, from the perspective 
of storing and extracting information. 

Unlike the wave function of a 'normal' atom, which carries 
the information just about its momentum, internal state and its binding energy, 
the wave function of a 'multi-site' atom contains in addition 
also the information about the parameters ($a$, $b$, $d$, $N_0$) 
of the diffraction grating. 

In particular, according to our results for the photo cross sections discussed above,  
the information about the grating, which is encoded in the quantum state of the atom,   
can be fully decoded by measuring the momentum distributions of 
the emitted electrons and the recoil ions. However, 
while the full information about the grating could be read off from 
the recoil ion spectrum alone, the electron momentum distributions 
can provide the information only if conditions are set 
on the momentum of the recoil ions and even in such a case 
it will not be complete (in general the number of slits $N_0$ 
cannot be extracted from the electron spectra due 
to the 'saturation' effect, see subsection A of this Section).

\subsection{ A 'multi-site' atom versus a multi-atomic molecule }

In our discussion of photo ionization of a 'multi-site' atom we 
stressed that clear (and profound) interference structures 
may arise in the spectra of emitted electrons and recoil ions 
and that they originate in a multi-site structure of the state 
of the atom passed through a diffraction grating.  

Interference effects can play a substantial role 
in photo ionization of molecules.   
In this case their origin is a multi-site structure 
of the electron state caused by the presence of more than one 
binding center (nuclei). 

The situation, in which an electron is simultaneously 
localized around more than one 'real' nuclei, is  
of course qualitatively different compared to that 
where a 'multiple localization' effectively arises due to       
alternating maxima and minima of the wave function of 
a single atom with just one 'real' nucleus. 
As a result, the interference 
effects in these cases will also qualitatively differ. 

\subsection{ Experimental devises as a possible source of decorrelation }
 
As we have seen, even in the single-slit case ($N_0=1$) the exact correlation between 
the momenta of the photo electrons and recoil ions is absent. However, a 'single-slit' case is 
essentially what one routinely encounters in experiments which use `nozzle/skimmer' or `collimator' devises.  
Therefore, their parameters should be taken into account when highly accurate experiments are performed.  
In particular, the necessity to do this was demonstrated in \cite{{proj-coherence }} 
where atom 
ionization by energetic charged projectiles was considered. 
 
\section{ Conclusions }
    
We have considered photo ionization (breakup) of an atom which 
passed through a diffraction grating. 
Due to scattering on the grating 
the atom finds itself in a state whose 
wave function possesses a periodic space structure 
with alternating maxima and minima. In the momentum space 
this results in the appearance of the alternating allowed 
and forbidden momentum bands.  

We have shown that such a multi-site structure of the atomic state 
qualitatively change the process of photo ionization 
compared to the case of a 'normal' atom. 
In particular, spectra of the emitted electrons and recoil ions 
demonstrate clear interference patterns arising due 
to one- or two-particle interference phenomena. 
This interference, however, has a qualitatively different character 
compared to that known to arise when molecules are photo ionized. 

Our results also suggest that the full information about 
the atomic properties and the diffraction grating, which is stored 
in the multi-site state of an atom, can be extracted 
in the process of photo ionization by measuring the momentum spectra of 
the emitted electrons and recoil ions. 
 
Unlike in photo ionization of a 'normal' atom or molecule, 
in the case of a 'multi-site' atom 
there is no mirroring between the momentum spectra of 
the photo electrons and recoil ions. In particular, 
the spectra of recoil ions display both one- and two-particle 
interference effects, which is not the case with 
the electron spectra where only the latter can be observed. 
Besides, the recoil-ion spectra contain also more information about 
the properties of the diffraction grating.    
The basic reason for these inequalities between the reaction 
fragments is a very large difference 
between the electron and nucleon masses which 
in the case of a 'multi-site' atom has a qualitatively different impact 
on the photo ionization process compared to the 'normal' case. 

\section*{Acknowledgement} 

We acknowledge the support from the National Key Research 
and Development Program of China (Grant No. 2017YFA0402300), 
the CAS President's International Fellowship Initiative  
and the German Research Foundation (DFG) under Grant No 349581371 
(the project VO 1278/4-1).


\begin{thebibliography}{99}  

\bibitem{coltrims} R. D\"orner, V. Mergel, O. Jagutzki, L. Spielberger, J. Ullrich, 
R. Moshammer and H. Schmidt-B\"ocking. Phys. Rep. {\bf 330}, 95-192 (2000); 
J. Ullrich, R. Moshammer, A. Dorn, R. D\"orner, L. Ph. H.
Schmidt, and H. Schmidt-B\"ocking, Rep. Prog. Phys. {\bf 66}, 1463-1545 (2003).

\bibitem{we-qdg} 
S. F. Zhang, B. Najjari, X. Ma and A. B. Voitkiv, 
arXiv:2002.09329 [quant-ph] (submitted). 

\bibitem{H-FP} M. Born and E. Wolf, {\it Principles of Optics} 
(Cambridge University Press, Cambridge, 1999).    

\bibitem{2-particle-2} 
J. W. Pan, Z.-B. Chen, C. Y. Lu, H. Weinfurter, A.
Zeilinger, and M. Zukowski, Rev. Mod. Phys. {\bf 84}, 777 (2012). 

\bibitem{2-particle-4} M. Waitz, D. Metz, J. Lower et al. 
Phys. Rev. Lett. {\bf 117}, 083002 (2016). 

\bibitem{proj-coherence } K. N. Egodapitiya,  S. Sharma,  A. Hasan,  A. C. Laforge,  D. H. Madison,  R. Moshammer,  and M. Schulz, Phys. Rev. Lett. {\bf 106}, 153202 (2011).  

\end{thebibliography}
\end{document}